\documentclass[12pt,preprint]{aastex}

\newcommand{\bdv}[1]{\mbox{\boldmath$#1$}}

\def\au{{\rm AU}}

\def\masyr{{\rm mas}\,{\rm yr}^{-1}}
\def\kpc{{\rm kpc}}
\def\mas{{\rm mas}}

\def\muas{\mu{\rm as}}

\def\max{{\rm max}}

\def\rel{{\rm rel}}

\def\e{{\rm E}}

\def\bmu{{\bdv\mu}}

\begin{document}
\title{Shortest Microlensing Event with a Bound Planet: KMT-2016-BLG-2605}

\author{\textsc{Yoon-Hyun Ryu$^{1}$, Kyu-Ha Hwang$^{1}$, Andrew
Gould$^{2,3}$, Jennifer C. Yee$^{4}$, Michael D.Albrow$^{5}$, Sun-Ju
Chung$^{1,6}$, Cheongho Han$^{7}$, Youn Kil Jung$^{1}$, Hyoun-Woo
Kim$^{1,8}$, In-Gu Shin$^{1}$, Yossi Shvartzvald$^{9}$, Weicheng
Zang$^{10}$, Sang-Mok Cha$^{1,11}$, Dong-Jin Kim$^{1}$, Seung-Lee
Kim$^{1,6}$, Chung-Uk Lee$^{1,6}$, Dong-Joo Lee$^{1}$, Yongseok
Lee$^{1,11}$, Byeong-Gon Park$^{1,6}$, Richard W. Pogge$^{3}$\\}}

\affil{$^{1}$Korea Astronomy and Space Science Institute, Daejon
34055, Republic of Korea}

\affil{$^{2}$Max-Planck-Institute for Astronomy, K\"{o}nigstuhl 17,
D-69117 Heidelberg, Germany}

\affil{$^{3}$Department of Astronomy, Ohio State University, 140 W.
18th Ave., Columbus, OH 43210, USA}

\affil{$^{4}$ Center for Astrophysics $|$ Harvard \& Smithsonian, 60
Garden St., Cambridge, MA 02138, USA} \affil{$^{4}$Department of
Astronomy and Space Science, Chungbuk National University, Cheongju
28644, Republic of Korea}

\affil{$^{5}$University of Canterbury, Department of Physics and
Astronomy, Private Bag 4800, Christchurch 8020, New Zealand}

\affil{$^{6}$Korea University of Science and Technology, Korea,
(UST), 217 Gajeong-ro, Yuseong-gu, Daejeon, 34113, Republic of
Korea}

\affil{$^{7}$Department of Physics, Chungbuk National University,
Cheongju 28644, Republic of Korea}

\affil{$^{8}$Department of Astronomy and Space Science, Chungbuk
National University, Cheongju 28644, Republic of Korea}

\affil{$^{9}$Department of Particle Physics and Astrophysics, Weizmann 
Institute of Science, Rehovot 76100, Israel}

\affil{$^{10}$Physics Department and Tsinghua Centre for
Astrophysics, Tsinghua University, Beijing 100084, China}

\affil{$^{11}$School of Space Research, Kyung Hee University,
Yongin, Kyeonggi 17104, Republic of Korea}

\begin{abstract}
KMT-2016-BLG-2605, with planet-host mass ratio $q=0.012\pm 0.001$,
 has the shortest Einstein timescale, $t_\e = 3.41\pm 0.13\,$days,
of any planetary microlensing event to date.  This prompts us
to examine the full sample of 7 short ($t_\e<7\,$day) planetary
events with good $q$ measurements.  We find that six have
clustered Einstein radii $\theta_\e = 115\pm 20\,\muas$ and
lens-source relative proper motions $\mu_\rel\simeq 9.5\pm 2.5\,\masyr$.
For the seventh, these two quantities could not be measured.
These distributions are consistent with a Galactic-bulge
population of very low-mass (VLM) hosts near the hydrogen-burning limit.
This conjecture could be verified by imaging at first adaptive-optics light
on next-generation (30m) telescopes.  Based
on a preliminary assessment of the sample, ``planetary'' companions
(i.e., below the deuterium-burning limit) are divided into
``genuine planets'', formed in their disks by core accretion,
and very low-mass brown dwarfs, which form like stars.  We discuss
techniques for expanding the sample, which include taking account
of the peculiar ``anomaly dominated'' morphology of the
KMT-2016-BLG-2605 light curve.

\end{abstract}

\keywords{gravitational lensing: micro}

\section{{Introduction}
\label{sec:intro}}

Microlensing planets are almost always discovered from short-lived
perturbations on otherwise single-lens single-source (1L1S) bell-shaped
\citet{pac86} light curves, as predicted by \citep{mao91}, \citet{gouldloeb},
and \citet{griest98}.
However, there are occasional exceptions.

If the planet-host mass ratio $q$ is relatively large and the
lens-source separation (normalized to the Einstein radius, $\theta_\e$)
is close to unity, $s\sim 1$, then the central and planetary
caustics merge into a single, large, resonant caustic, which can
induce a long-term anomaly over the peak of the event.  For example,
the $q=10^{-2}$ event MOA-2009-BLG-387 \citep{mb09387} showed strong,
continuous anomalies over 9 days.  Nevertheless, over the remainder
of the $2\,t_\e\sim 80\,$days of the event, it appeared as qualitatively
normal.  And, indeed, it is on the basis of this normal rising
behavior that the Microlensing Follow Up Network ($\mu$FUN)
initiated followup observations two days before the anomaly.  Here,
$t_\e$ is the Einstein timescale.

A very different counter-example, in this case a ``purely anomalous event''
is given by MOA-bin-1 \citep{moabin1}, for which the entire observed
event consists of the source crossing the planetary caustic.  Because
the planet-host separation is $s\sim 2$ and the source trajectory is
nearly perpendicular to the planet-host axis, $\alpha\sim 90^\circ$,
the $t_\e\sim 31\,$day ``host event'' leaves barely a trace on the
light curve.  Thus, essentially all that remains is the short, $\sim 0.2\,$
day triangular anomaly due to the $q\sim 5\times 10^{-3}$ planet.

Here we present another rare case for which the light curve is dominated
by a planet-induced anomaly, KMT-2016-BLG-2605.  Like MOA-2009-BLG-387,
the anomaly is due to a resonant caustic of a $q\sim 10^{-2}$ planet.
And like MOA-bin-1, the duration of the anomaly is short ($1.5\,$days).
Indeed, the observed portion of the anomaly is only $\sim 0.5\,$ days.
However, in contrast to either of these cases, the underlying timescale
is very short, $t_\e=3.4\,$days, while the microlensed source is faint,
$I_s=20.2$, so that only the anomaly is really noticeable, particularly
in the initial reductions from which the event was discovered.  For
this reason, the event was not discovered in the original search
carried out by the Korean Microlensing Telescope Network
(KMTNet, \citealt{kmtnet}) EventFinder system \citep{eventfinder},
during which it was misclassified as a cataclysmic variable (CV).
It was recognized as genuine microlensing only as a byproduct
of a special search that was conducted for another purpose \citep{kb192073}.

KMT-2016-BLG-2605 is one of only seven short-timescale ($t_\e<7\,$day)
planetary events with a well measured mass ratio $q$ (less than
factor 2 difference between competing solutions at $\Delta\chi^2<10$).
The roughly comparable properties of this ensemble
($\theta_\e\sim 0.11\,\mas$, $\mu_\rel \sim 9\,\masyr$,
where $\mu_\rel$ is the lens-source relative proper motion) are
consistent with a population of Galactic-bulge hosts that have masses near the
hydrogen-burning limit.  For six of these seven
(including KMT-2016-BLG-2605), this
assessment can be confirmed or contradicted at first
adaptive-optics (AO) light on next-generation (30m class) telescopes.

Despite the fact that all seven were detected through
resonant or near-resonant caustics, KMT-2016-BLG-2605 is the only
anomaly-dominated event, which would potentially make it more
difficult to recognize during the manual stage of event selection.
We consider this and other factors to evaluate the
challenges in identifying more of these short-$t_\e$ planetary systems.

\section{{Event Identification}
\label{sec:identify}}

KMT-2016-BLG-2605 was identified as a ``clear'' microlensing event
during a special search of the 2016 KMTNet data that was conducted
as part of a project to identify all finite-source point-lens (FSPL)
giant-source events during the four year span 2016-2019.  The primary
objective of this project is to create a statistically well defined parent
sample that will contain a free-floating planet (FFP) subsample.
A special additional search was found to be necessary because
a significant fraction of short FSPL events were missed by the
primary EventFinder \citep{eventfinder} searches that are done annually.
For example, some giant-star sources had been eliminated from the
search because of previously cataloged variability, or light-curve
artifacts.  Of course, it is known that variable stars can undergo microlensing
events that can be distinguished from their intrinsic variations,
and also that many light-curve artifacts do not repeat from year to year.
However, the fraction of real events that are removed this way is small
$(\sim 1\%)$, while the human cost of manual review is high.  Thus,
removal of these sources from the regular search is a rational approach.
Nevertheless, when the search is restricted to giant sources, the cost
is reduced by a factor of about 20, making lifting (or strongly
modifying) these criteria worthwhile for the FSPL project.

Another feature of the special search was that the candidates
selected by the machine algorithm were shown to the operator in
several additional displays.  This is because the shortest FSPL
events can be highly anomalous due to finite source effects.
Thus, when the three data sets are automatically aligned using
\citet{pac86} or \citet{gould96} fits, the joint light curve can
appear to be ``clearly not microlensing''.  By having multiple
displays, by having the operator spend more time reviewing
each candidate, and by adopting somewhat lower standards
on what is a plausible microlensing event, it is much less likely that these
short FSPL events will be rejected at this stage.  Of course,
non-microlensing events can still be rejected at a later stage when each
candidate is manually fit to point-source point-lens (PSPL) and FSPL
functional forms.  For more details on these special searches,
see \citet{kb192073} and \citet{kb172820}.

The special search for 2016 identified 281 candidates, of which 37
had not been found in the regular EventFinder search\footnote{Of these
37, 10 had previously been identified by other teams, including
seven by OGLE, two by MOA and one by both.}.
KMT-2016-BLG-2605 was number 17 on this list of 37 new candidates.
Following the convention of \citet{ob161928} and \citet{kb172820},
it was assigned the sequential label ``2605'' ($=2588 + 17$)
because there were 2588 events discovered in the original search.

KMT-2016-BLG-2605 is neither an FSPL event nor does it have
a giant-star source.  Nevertheless, it passed the various
selection criteria imposed to obtain a sample of just
seven new events\footnote{Two of these seven proved to be FSPL events.}
from 2016 that would then be subjected to manual FSPL fitting.
The machine search of the KMTNet database is restricted to
``giant'' source stars, defined
as having dereddened baseline magnitudes
$I_{\rm base,0} = I_{\rm cat} - A_I< 16.2$, where $I_{\rm cat}$ is the
magnitude of the catalog entry,  $A_I= A_K/7$, and $A_K$ is derived
from \citet{groenewegen04}.  This led to $I_{\rm base,0} = 18.63 - 2.74 = 15.89$.
Being a very short event, it then easily passed the machine-search
criterion that the effective timescale be less than 5 days.  When it
was displayed to the operator, it did not look like 1L1S microlensing
(neither PSPL nor FSPL), but at this stage the only criterion is
that the variation is plausibly due to microlensing.  As described
by \citet{kb192073}, all events (including these 37 from the 2016 special
search) are selected for manual review by two criteria.  The first
is $I_{s,0} = I_s - A_I<16$, where $I_s$ is the source magnitude from
the pipeline fit to the event.  As we will show in Section~\ref{sec:anal},
the true value of $I_{s,\rm kmt} = 20.0$, for which case $I_{s,0} = 17.3$, i.e.,
failing this criterion by more than a magnitude.  Nevertheless, due to
the extremely anomalous form of the light curve, the pipeline fit assigned
$I_s \rightarrow I_{\rm cat}$, which allowed the event to pass this criterion.
Second, it easily passed the criterion meant to select plausibly FSPL
(as opposed to almost certainly PSPL) events,
$\mu_{\rm thresh} = 3\,\mas\, 10^{(16-I_{s,0})/5}/u_0 t_\e > 1\,\masyr$,
where $u_0$ is the impact parameter (normalized to $\theta_\e$)
of the pipeline fit.  Given that
$u_0 t_\e= 0.147\times 1.59\,{\rm day} = 5.6\,$hr, it had
$\mu_{\rm thresh} = 5\,\masyr$.

It was only in the course of
fitting the event to 1L1S models by hand that it became clear
that the slope of the light curve showed a discontinuous change
at HJD$^\prime ={\rm HJD}-2450000 = 7565.3$, indicating a caustic
crossing, which could plausibly be explained by a planetary system.

In brief, KMT-2016-BLG-2605 came to our attention by a most
unlikely and circuitous path, a point to which we will return
in Section~\ref{sec:discuss}.  For example, if the source
star were not blended with another star that was several times brighter,
it would not have even been selected for machine fitting at the
first step.

\section{{Observations}
\label{sec:obs}}

KMT-2016-BLG-2605 is at equatorial coordinates
(R.A.,Decl)$_{J2000} = $(17:59:17.54, $-$26:58:55.20), corresponding
to galactic coordinates $(l,b) = (+3.22,-1.60)$.  It therefore
lies in KMTNet field BLG03.  KMTNet consists of three 1.6m telescopes,
each equipped with a $4\,{\rm deg^2}$ camera, and located in
Chile (KMTC), South Africa (KMTS), and Australia (KMTA).  At the time
of the event, BLG03 was observed with a cadence $\Gamma=2\,{\rm hr}^{-1}$
from each observatory, primarily in the $I$ band.
In 2016, every tenth $I$-band observation at KMTC was complemented
by one in the $V$ band, as was every twentieth observation at KMTS.
There were no $V$-band observations from
KMTA\footnote{Based on experience with OGLE-2015-BLG-1459 \citep{ob151459},
it was realized that KMTA $V$-band observations were potentially very
important, and ultimately both KMTS and KMTA were observed in
the $I$ and $V$ bands  at a ratio of $10:1$.
Unfortunately, this was not the case in 2016.}.

The data were initially reduced using pySIS \citep{albrow09},
which is a specific implementation of difference image
analysis \citep{tomaney96,alard98}.  Moreover, the original
light-curve analysis was based on tender loving care (TLC)
pySIS re-reductions.  However, as we describe in Section~\ref{sec:anal},
it was ultimately necessary to use a related package, pyDIA \citep{pydia}
because it returns field-star photometry on the same system as the
light curve.
To avoid confusion, we present the entire investigation of the
light curve using pyDIA photometry.


\section{{Light Curve Analysis}
\label{sec:anal}}

We fit the data to binary-lens single-source (2L1S) models, which
are characterized by seven parameters $(t_0,u_0,t_\e,s,q,\alpha,\rho)$,
where $t_0$ is the time of closest lens-source approach and
$\rho=\theta_*/\theta_\e$ is the source radius normalized to the
Einstein radius.  As is almost always done, we begin with a grid search
over an $(s,q,\alpha)$ grid, in which $(s,q)$ are held fixed and
$(t_0,u_0,t_\e,\alpha,\rho)$ are allowed to vary in a Monte Carlo Markov chain
(MCMC). The three \citet{pac86}  parameters $(t_0,u_0,t_\e)$ are seeded
at the PSPL fit, while $\rho$ is seeded at $\rho=0.003$.

\begin{figure}
\plotone{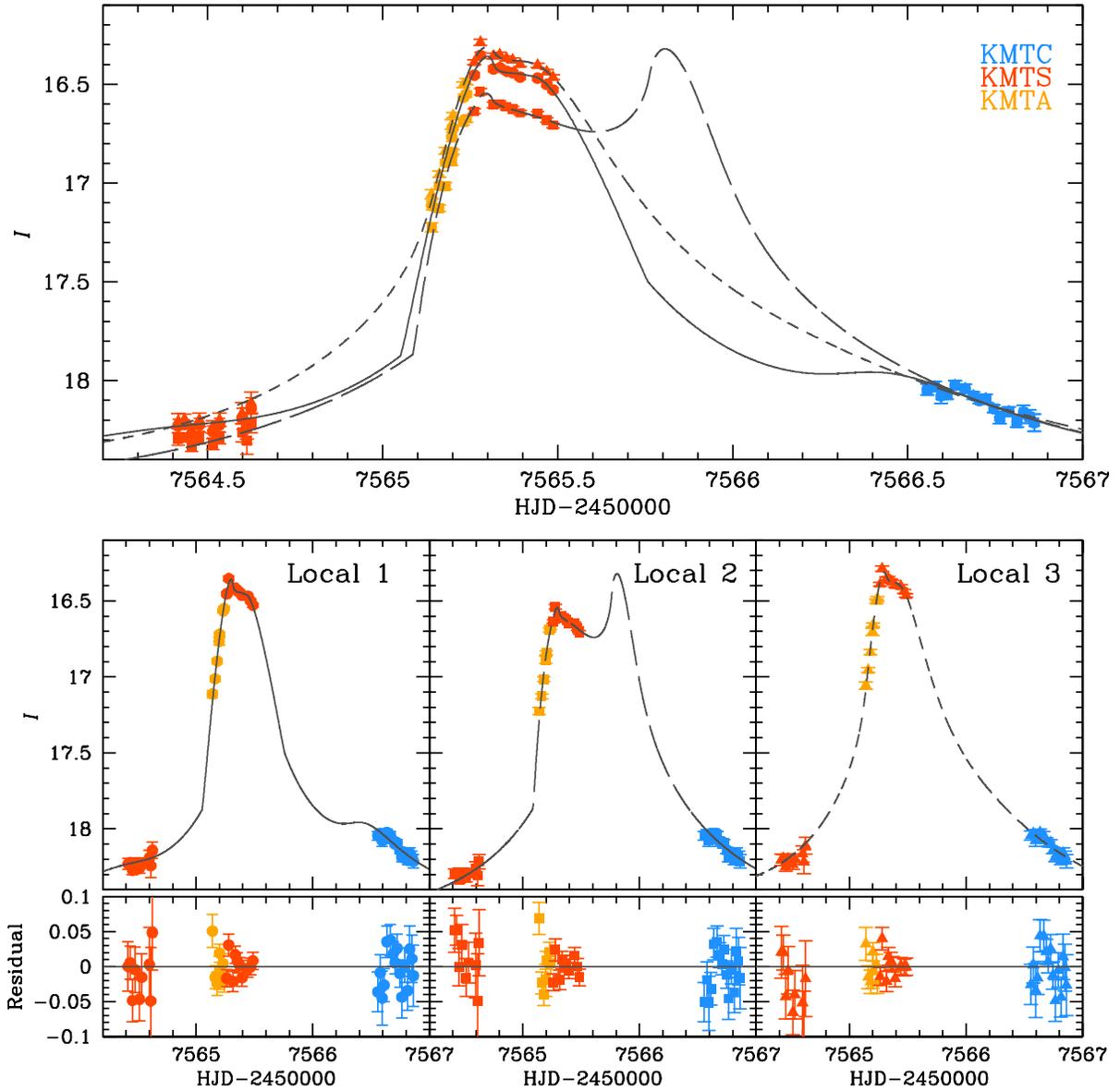}
\caption{Light curve and models for Locals~1--3 for KMT-2016-BLG-2605.
Upper Panel: In contrast to the great majority of microlensing events, the data
points for competing models are offset from one another for both of the
non-anchor observatories (KMTS and KMTA).  By construction, they are perfectly
aligned for the anchor (flux-reference) observatory (KMTC).
These offsets occur because each observatory
has only one night of strongly magnified data.  See Section~\ref{sec:identify}.
The middle and bottom rows of panels show the individual models and their
residuals, respectively.  By eye, Local~1 better matches the data.  Formally,
the fit is better by $\Delta\chi^2\ga 10$.  See Table~\ref{tab:local_1-3}.}
\label{fig:lc1-3}
\end{figure}

\begin{figure}
\plotone{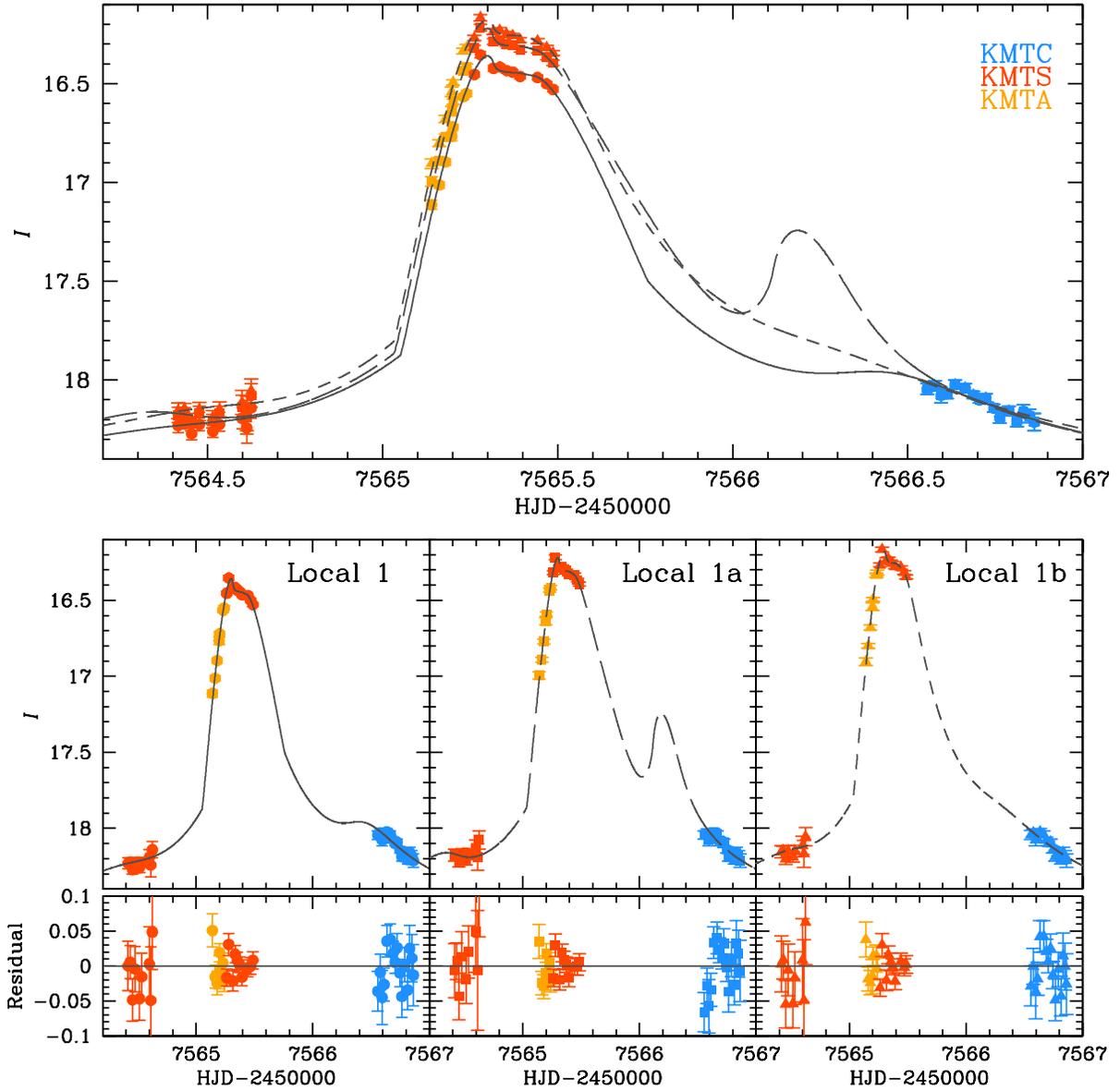}
\caption{Similar to Figure~\ref{fig:lc1-3}, but for Local~1 and its two
satellite solutions, Local~1a, and Local~1b.  By eye, Local~1 is favored
over the other two, but less decisively than for Figure~\ref{fig:lc1-3}.
See Table~\ref{tab:local_1-3}.}
\label{fig:lc1ab}
\end{figure}

However, in contrast to typical experience, we find a plethora of
quite distinct solutions.  After refitting each local minimum of the
grid with all seven parameters allowed to vary, we find seven
distinct solutions within the range of $\Delta\chi^2<100$.
However, only Locals 1--3 are potentially viable, with
$\Delta\chi^2 < 11$, while Locals 4--7 have $\Delta\chi^2 > 50$.
Closer examination of Local 1 shows that it breaks up into
three nearby minima.  See Table~\ref{tab:local_1-3}.
Figure~\ref{fig:lc1-3} shows the models and data for the
three principal Locals, while Figure~\ref{fig:lc1ab} shows the
principal Local 1 and its two satellite solutions.


Comparing Locals~1 and 1b in Figure~\ref{fig:lc1ab}, 
we see that the
peak of the former (specifically, the second KMTS point at
HJD$^\prime = 7565.28$) is 0.25 mag fainter.  This is
unusual for competing microlensing solutions.  When presenting microlensing
models, one data set (in this case, KMTC) is chosen as the ``anchor''.
Its data values are exactly reproduced in the figure.  All other data
sets are aligned, by linear regression of the fluxes to the model,
to this anchor.  Because this alignment is usually based on several
nights of data during which the event is evolving in a regular way,
the alignment coefficients are normally the same for different models.
Hence, the different data sets are rigidly aligned to the same
fiducial scale, independent of model.  For this reason, all models
can be shown by curves that are superposed on data points whose positions
are fixed.

However, in the present case, the KMTS and KMTA data are strongly
magnified on only one night.  Hence the alignment is not fully
constrained by other nights.  The lack of a rigid constraints
is reflected in the range of values in the
quantity $\Delta I=I_{S,,\rm kmts}-I_{S,\rm kmtc}$ that is shown
in Table~\ref{tab:local_1-3}.

At most one of these $\Delta I$ values can be correct.
That is, these offsets represent
the relative transparency and throughput of the detectors at the
two sites.  And this quantity can be measured from field stars
\citep{gould10,mb11293}.

To make this comparison precise, we have carried out the
MCMCs  with the source-flux and blend-flux parameters from each observatory
treated as chain variables,  That is, normally one writes
\begin{equation}
F_{i}(t_{i,j}) = f_{s,i}A(t_{i,j};t_0,u_u,t_\e,s,q,\alpha,\rho) + f_{b,i}\ ,
\label{eqn:fsfb}
\end{equation}
where $F_i(t)$ is the observed flux from observatory $i$ at time $t$,
and $(f_{s,i},f_{b,i})$ are the source-flux and blend-flux parameters
for observatory $i$.  At each step on the MCMC, one inserts the
trial values for the seven parameters $(t_0,u_u,t_\e,s,q,\alpha,\rho)$,
but one determines $(f_{s,i},f_{b,i})$ from a linear fit to the model
magnifications $A$.  For the vast majority of cases, the errors
that would be induced in these parameters due to the flux errors at
a given model is tiny compared their error due to variation between
different models.  Hence, this approach is usually appropriate.

However, for the present case, the value of $f_{s,\rm kmtc}$ is basically
determined by just 16 data points on the night after the peak.  These
individually have fractional scatter $\sigma(F)/[(A-1)f_s\sim 0.075$,
which implies that the standard error of the mean should be of order
$(2.5/\ln 10)\times 0.075/\sqrt{16} \sim 0.02$ mag.  This is too large
to be ignored in the present context.  Hence, we also treat $f_s$ and
$f_b$ as chain parameters.

\begin{figure}
\plotone{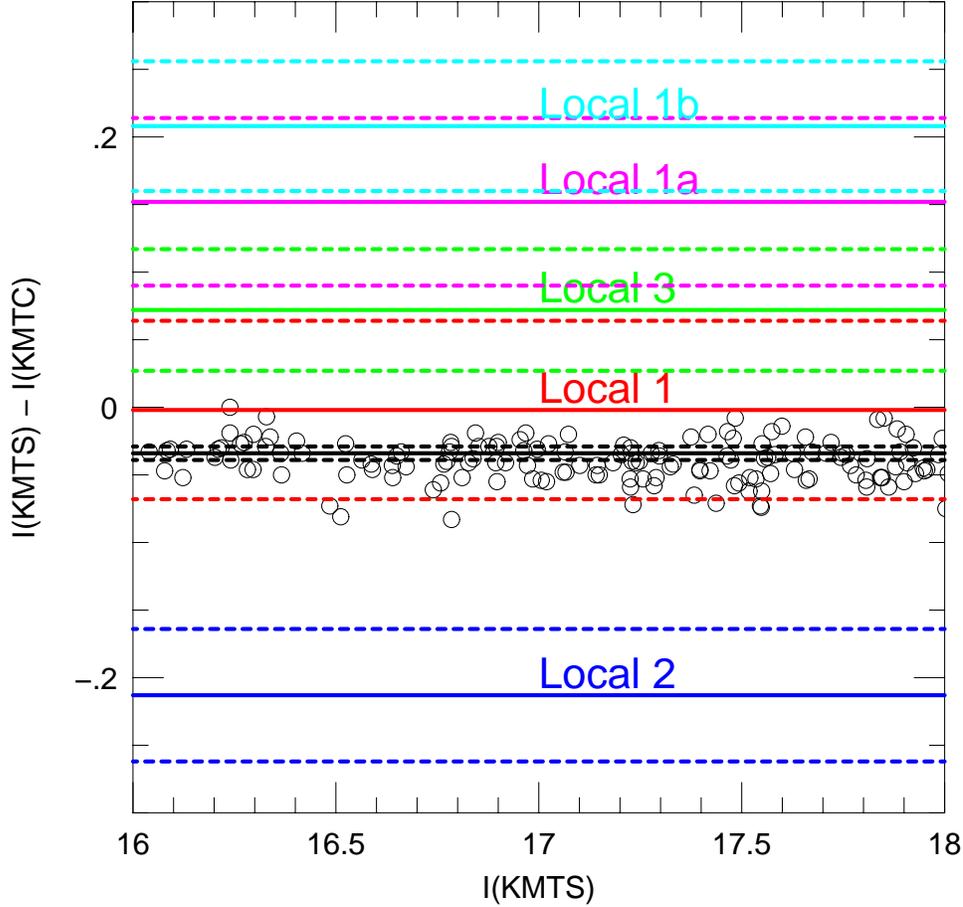}
\caption{Offset between KMTS and KMTC $I$-band photometry as determined from
field stars (open circles), compared to the predictions of the five different
models shown in Figures~\ref{fig:lc1-3} and \ref{fig:lc1ab}. See the penultimate
row of Table~\ref{tab:local_1-3}.  There is $1\,\sigma$ agreement for Local~1
and $\geq .3\,\sigma$ disagreement for all the others.  After incorporating
the flux constraint (black dashed band) into the MCMC, Locals~1a and 1b are
eliminated as distinct minima, while Locals~2 and 3 become disfavored by
$\Delta\chi^2\ga 16$.  See Table~\ref{tab:const_1-3}.
}
\label{fig:offset}
\end{figure}

Finally, we match the field-star photometry of the KMTS03 and KMTC03
reductions, and we plot the differences as a function of magnitude
in Figure~\ref{fig:offset}.  For this purpose, we only include stars
with $(V-I)_{\rm kmts}>3.0$, which includes the colors of the source
and the red clump.  We plot the predicted offsets from the five
models as horizontal lines with error ranges.

We see that $(I_{s,\rm kmts} - I_{s,\rm kmtc})_{\rm Local-1}$ is consistent with
the field stars at $1\,\sigma$, whereas $(I_{s,\rm kmts} - I_{s,\rm kmtc})$
is inconsistent at $\geq 2.3\,\sigma$ for all the other solutions.
Keeping in mind that Local~1 was already favored over Locals~2 and 3
by $\Delta\chi^2\ga 10$ (and by slightly less compared to its satellite
solutions), we regard this as clear confirmation of Local~1.

\begin{figure}
\plotone{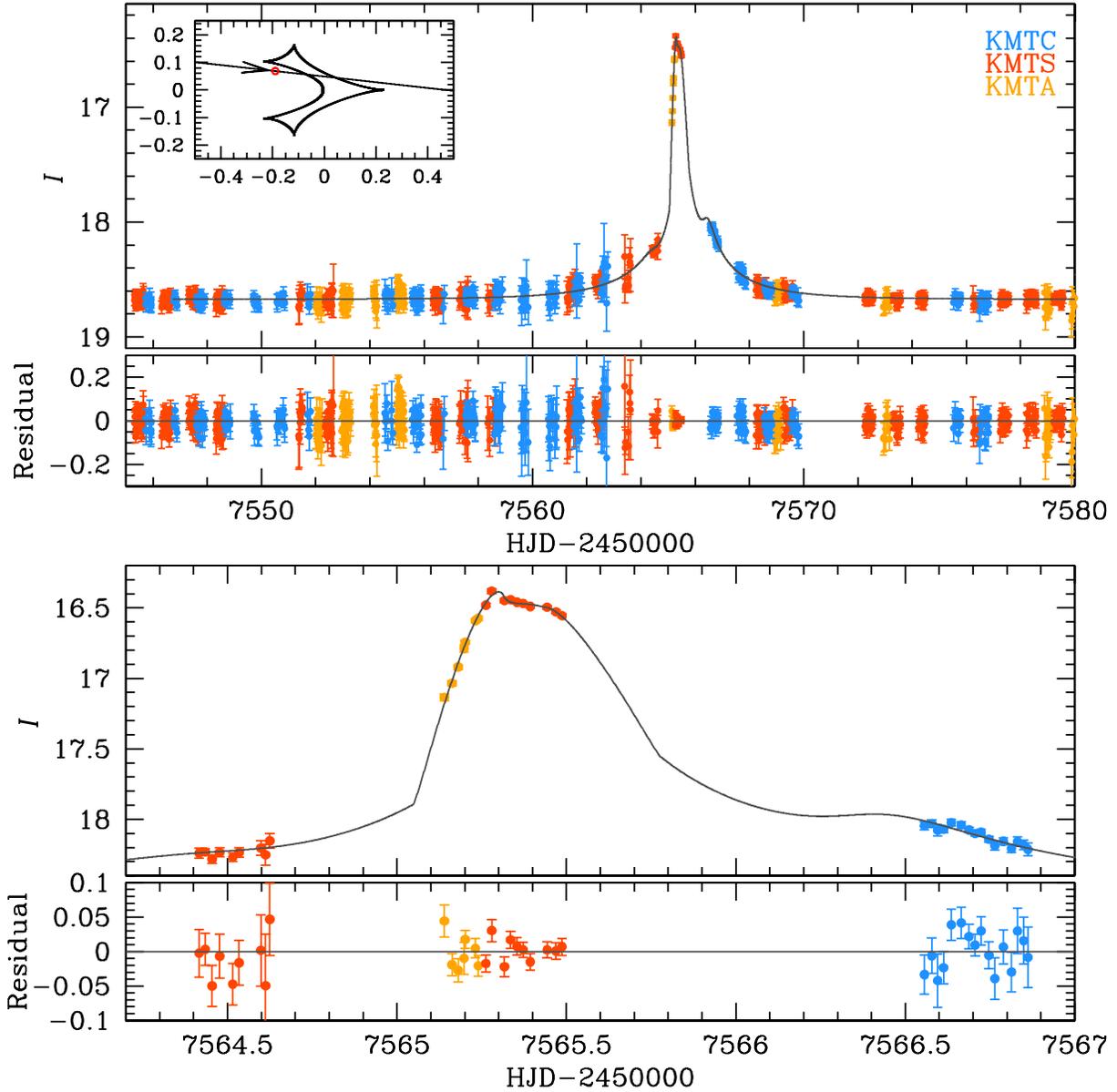}
\caption{Light curve and models for Local~1 for KMT-2016-BLG-2605, after
incorporating the flux constraint shown in Figure~\ref{fig:offset}.
Also shown is the caustic topology for the event.
}
\label{fig:lc}
\end{figure}

Finally, we impose the flux constraint within the MCMC as a $\chi^2$
penalty,
\begin{equation}
\chi^2_{\rm flux} = \biggl({-2.5\log(f_{S,\rm kmts}/f_{S,\rm kmtc}) - -0.033
\over 0.005}\biggr)^2 .
\label{eqn:chi2pen}
\end{equation}
We find that Locals~1a and 1b disappear as separate minima,
while Locals~2 and 3 are each disfavored by $\Delta\chi^2\sim 16$.
See Table~\ref{tab:const_1-3}.
We adopt the Local~1  microlensing parameters in this table as our final
result.  Figure~\ref{fig:lc} shows the best-fit model and data after
imposing this constraint. It also shows the caustic topology, which
is resonant.

\section{{Source Properties}
\label{sec:cmd}}

As with most other microlensing events, we measure $\theta_*$
using the method of \citet{ob03262}.  This requires that
we first find
the offset $\Delta[(V-I),I] = [(V-I),I]_s - [(V-I),I]_{\rm cl}$ of
the source star relative to the clump.  Adopting
 $[(V-I),I]_{\rm cl,0} = (1.06,14.34)$ \citep{bensby13,nataf13},
we would then derive the dereddened source color and magnitude,
$[(V-I),I]_{s,0} = \Delta [(V-I),I] + [(V-I),I]_{\rm cl,0}$, convert
from $V/I$ to $V/K$ photometry using the $VIK$ color-color relations
of \citet{bb88}, and finally, use the color/surface-brightness
relation of \citet{kervella04} to derive the angular radius of the
source star, $\theta_*$.

Unfortunately, the first step in this procedure, determining
$\Delta[(V-I),I]$, poses substantially greater challenges for
KMT-2016-BLG-2605 than it does for typical events because
the color-magnitude diagram (CMD) positions of both the source star
and the clump are more difficult to determine.

In the Appendix, we delineate the steps to measure the clump position,
which we show is well determined.  However, we show that while the source
magnitude is also well determined, the source color remains somewhat
ambiguous.  The key issue is that the source color derived from the
light curve, which rests on a single magnified $V$-band data point,
is in formal conflict with Bayesian expectations based on the well-determined
magnitude (together with the morphology of the CMD).  After weighing
all the evidence, we conclude that, most likely, the discrepancy
is due to a relatively large $(\la 3\,\sigma)$
statistical error in the single $V$-band
data point, and we adopt
\begin{equation}
(V-I)_{s,0} = 1.00\pm 0.05
\label{eqn:vmi-adopt1}
\end{equation}

Then, following the steps outlined in the first paragraph of this
section, we find
\begin{equation}
\theta_* = 1.38\pm 0.10\,\muas,
\label{eqn:thetastar}
\end{equation}
where we have added 5\% in quadrature to the error bar to take
account of systematics that are inherent to the method.
From this we then derive,
\begin{equation}
\theta_\e = {\theta_*\over \rho} = 0.116 \pm 0.009\,\mas;
\qquad
\mu_\rel = {\theta_*\over t_*} = 12.3\pm 1.0\,\masyr,
\label{eqn:thetae_murel}
\end{equation}
where $t_*\equiv \rho t_\e$.

However, in the Appendix, we also keep track of the possibility
that the light-curve color measurement is actually correct, in which case,
$\theta_* = 1.77 \pm 0.23\,\muas$,
$\theta_\e =  1.49\pm 0.19\,\mas$, and
$\mu_\rel =  15.8\pm \,1.3\,\masyr$.

\section{{Physical Parameters}
\label{sec:physical}}

\begin{figure}
\plotone{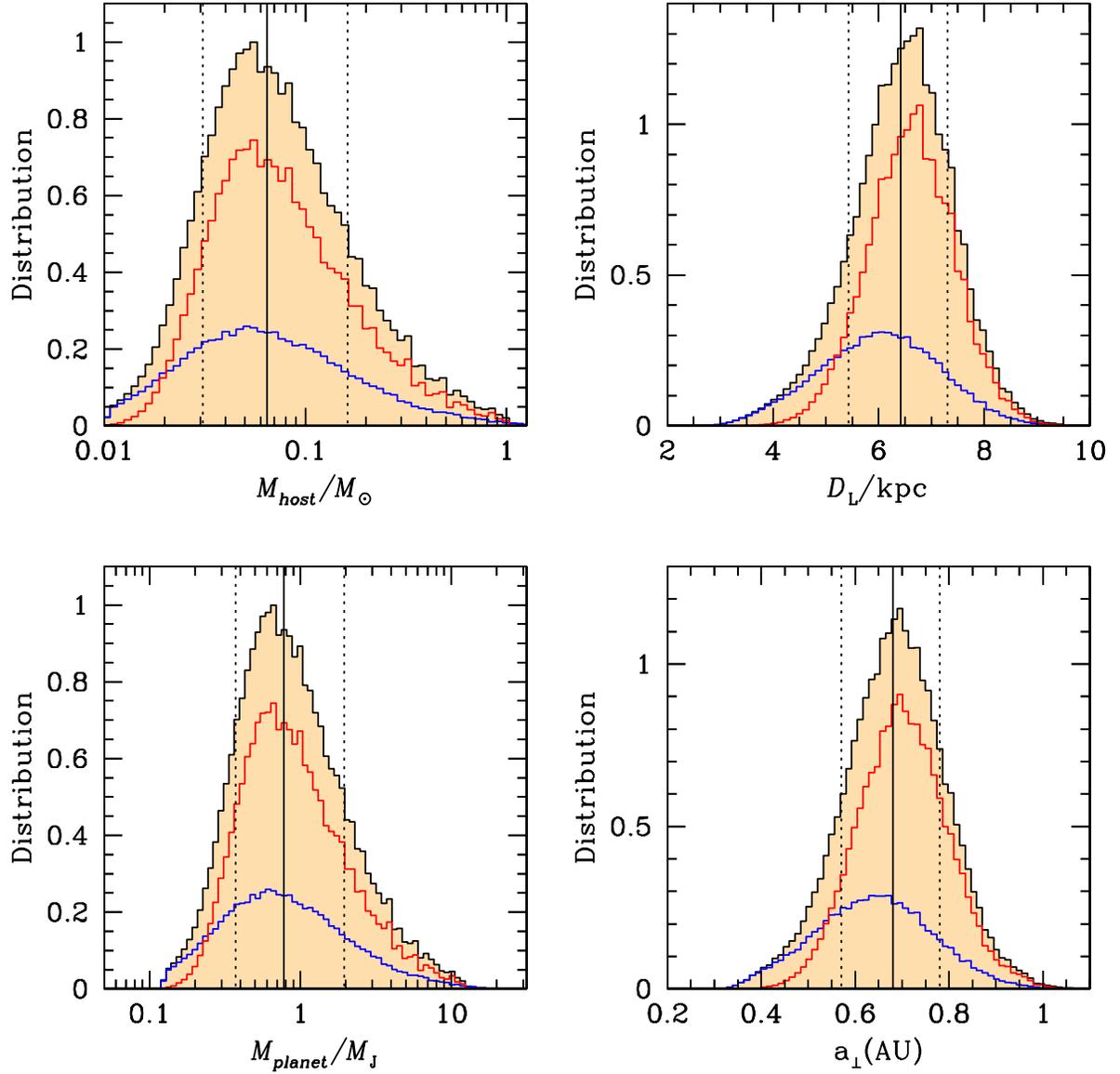}
\caption{Bayesian estimates of the host mass, planet mass, system distance,
and planet-host projected separation for KMT-2016-BLG-2605.  The red and
blue histograms show the relative contributions of bulge and disk lenses,
respectively, with the total are shown as black histograms.  The median host
mass is very close to the star-BD boundary.
}
\label{fig:bayes}
\end{figure}

We make a standard Bayesian analysis to derive physical parameters.
That is, we draw events randomly from a Galactic model,
and we weight each simulated event by how well it
conforms to Equation~(\ref{eqn:thetae_murel}).  We additionally weight
by the event rate, $\Gamma\propto \theta_\e\mu_\rel$, although this
has very little effect because these parameters are very similar
for all simulated events that satisfy Equation~(\ref{eqn:thetae_murel}).
The Galactic model follows that of \citet{ob171522} as modified by
\citet{ob180567}.
The results are reported in Table~\ref{tab:bayes}, and they are
illustrated in Figure~\ref{fig:bayes}, which shows that the lens
probably lies in the bulge, but even if not, it most likely lies in
overlapping regions of the disk.  The median host mass estimate,
$M_{\rm host} = 0.064_{-0.032}^{+0.099}$ is very close to the hydrogen-burning
limit, i.e., there is a roughly 50\% probability that it is a brown dwarf (BD).
If we adopt a snow-line scaling $a_{\rm snow} = 2.7\au (M_{\rm host}/M_\odot)$,
then the planets projected separation is $a_\perp \sim 4\,a_{\rm snow}$.

\citet{kb190371} have shown that, unless $\mu_\rel>10\,\masyr$
(or there is additional information such as a microlens parallax
measurement), the Bayesian mass estimate depends only on $\theta_\e$.
While this condition does not apply to KMT-2016-BLG-2605, if we nevertheless
input $\theta_\e=0.116\,\mas$ into their Figures 7, we obtain
$M = 0.08_{-0.04}^{+0.06}\,M_\odot$.  The difference between this and our
Bayesian estimate is accounted for by the ``bend'' in the median
trajectories above $\mu=10\,\masyr$ in their Figure 6.


\section{{Discussion}
\label{sec:discuss}}

\subsection{{Ensemble of Short-timescale Planets}
\label{sec:ensemble}}

At $t_\e=3.41\,$days, KMT-2016-BLG-2605 has the shortest timescale of
any planetary microlensing event.  There are eight previous binary-lens
microlensing events with $3.7 <t_\e/{\rm day}<7$ that are listed
by the NASA Exoplanet
Archive\footnote{https://exoplanetarchive.ipac.caltech.edu/.
We chose ``7 days'' as the upper limit, without foreknowledge
of the sample that it would produce, because 7 is the closest integer
to $2\times t_\e/$day of KMT-2016-BLG-2605.}.
For the present discussion, we restrict
attention to the subset with (1) unambiguous measurement of $q$
(specifically, no solutions with $\Delta\chi^2<10$ and $q$ values differing
by a factor $>2$); and (2) ``verifiable planet'', specifically
$q<M_{\rm D-burn}/M_{\rm H-burn}= 0.16$.  That is, we accept the formal
definition of a ``planet'' as having mass $m_p$ below the deuterium burning
limit, $m_p<M_{\rm D-burn}$.  Systems with $q$ above this limit can be ruled
out as planets if their hosts are stars $M_{\rm host}>M_{\rm H-burn}$
because these can be imaged at late times.  But a non-detection would
leave the status of the companion ambiguous.  However, for systems
satisfying this condition, even a non-detection would prove the
companion had planetary mass.
Two events fail criterion (1):
MOA-bin-29 \citep{moabin29} and
MOA-2015-BLG-337 \citep{mb15337}.
One event fails criterion (2): KMT-2016-BLG-2124 \citep{kb161820}.

The remaining five events are
MOA-2011-BLG-262 \citep{mb11262},
OGLE-2015-BLG-1771 \citep{ob151771}, 
KMT-2018-BLG-0748 \citep{kb180748}, 
OGLE-2018-BLG-0677 \citep{ob180677}, and 
KMT-2016-BLG-1820 \citep{kb161820}. 
We note that, strictly speaking, the first of these
events has an ambiguous measurement of $\rho$, i.e.,
$\rho=3.44\times 10^{-3}$ or $\rho=5.73\times 10^{-3}$, with the first
preferred by $\Delta\chi^2=3$.  However, the first solution would
imply a geocentric proper motion\footnote{The authors quote
$19.6\pm 1.6\,\masyr$, which may reflect a posterior result after
applying unstated Bayesian priors.} $\mu_\rel=21.6\pm 2.3\,\masyr$.
As the authors
note, their OGLE-III-based measurement of the source proper motion
$\bmu_s(l,b) \sim (-2.4,-0.4)\pm (2.7,2.7)\,\masyr$ implies that the
high proper-motion solution is inconsistent with bulge lenses.
It would imply
that the lens must lie far in the foreground, e.g., at
$D_L\la 1\,\kpc$.  In this case, the host mass would be $M\la 6\,M_J$,
with ``planet'' (aka ``moon'') mass $m\la 0.9\,M_\oplus$.  The ``novelty''
of this putative system, combined with the small number of potential
lenses in the nearby observational cone, renders this solution highly
unlikely.  Therefore, for this purpose, we adopt the higher-$\rho$ solution.

To this sample, we add KMT-BLG-2019-BLG-0371 \citep{kb190371}, which
is not listed at NASA Exoplanet Archive because it has not yet
been accepted for publication.

Table~\ref{tab:6event} shows the observed characteristics of the
six previous systems\footnote{The values in this table have
been somewhat compressed and simplified in order to aid visual
assimilation of the patterns.  The reader should consult the
original papers for the exact parameter values and error bars.},
together with those of KMT-2016-BLG-2605.
Excluding for the moment OGLE-2018-BLG-0677 (for which $\rho$ is not
measured), the remaining six events all have Einstein radii $\theta_\e$
in the range $115\pm 20\,\muas$, and three have proper motions
$\mu_\rel \sim 9.5\,\masyr$, with the other three deviating by 2--3 $\masyr$.

These characteristics are consistent with expectations for a population
of planet-bearing hosts near the star-BD boundary and lying in the
Galactic bulge.  That is, the total mass of these systems (dominated by
the host) is given by
\begin{equation}
M = {\theta_\e^2\over\kappa\pi_\rel} =
0.10\,M_\odot \biggl({\theta_\e \over 115\,\muas}\biggr)^2
\biggl({\pi_\rel \over 16\,\muas}\biggr)^{-1};
\qquad
\kappa\equiv {4 G\over c^2\au}\simeq 8.14\,{\mas\over M_\odot},
\label{eqn:massrange}
\end{equation}
where we have scaled to a typical relative parallax, $\pi_\rel = 16\,\muas$,
for bulge-bulge microlensing.  Under conditions that the ``mass function''
(i.e., in this case, the mass function of stars/BDs that host planets)
has a ``hard floor'', the shortest-timescale events will be generated
by systems near this floor, and with source-lens separations $D_{LS}$
near the ``edge'' of the bulge distribution.  In fact, there is no
real ``edge'', but there is a rapid fall-off.  It is also unlikely that
there a ``hard floor'' to the mass function, but it is plausible that
there is, again, a rapid fall-off.

The same picture naturally explains the high proper motions.  For
an isotropic proper-motion distribution with Gaussian width
$\sigma=2.9\,\masyr$ (which approximately characterizes the bulge),
the mean and standard deviation of the lens-source relative proper motion
is \citep{gould21},
\begin{equation}
\mu_\rel = \biggl({4\over\sqrt{\pi}} \pm \sqrt{6 - {16\over\pi}}\biggr)\sigma
\rightarrow 6.5\pm 2.8\, \masyr .
\label{eqn:pmdist}
\end{equation}
Thus, we expect that if there is a ``floor'' on the mass function
(whether ``hard'' or ``soft''), the proper motions of the
shortest events will tend toward
the upper range where the distribution is falling off rapidly,
roughly $1\,\sigma$ above the mean, which is $9.3\,\masyr$ in the
present case.

These are just plausibility arguments, and no more is really
possible at this point because of the inhomogeneous selection of
the sample.  However, it will be straightforward to test this
conjecture by imaging the systems at first AO light
on next-generation (30m class) telescopes, in roughly 2030.  In all
cases, the sources are dwarf stars, turnoff stars or subgiants, and hence
have $M_K\ga 2$, compared to $M_K\sim 10$ of stars at the bottom
of the main sequence, i.e., contrast ratios of $\la 8$ magnitudes.
\citet{bowler15} achieved contrast ratios of (5, 10) magnitudes at
separation $\Delta\theta\sim (150,320)\,\mas$
using AO on the Keck 10m telescope.  Scaling to 25m (for the Giant
Magellan Telescope, GMT) to 30m (for the Thirty Meter Telescope, TMT),
and to 39m (for the European Extremely Large Telescope, EELT),
these correspond to
$\Delta\theta\sim (60,130)\,\mas$,
$\Delta\theta\sim (50,110)\,\mas$, and
$\Delta\theta\sim (40,80)\,\mas$, respectively.  All but the last
two events in Table~\ref{tab:6event} will have $\Delta\theta\ga 110\,\mas$
by 2030, making them accessible down to the hydrogen burning limit
at either TMT or EELT, with a few requiring several additional years
for access from GMT.  KMT-2019-BLG-0371 would only be fully accessible
from EELT in 2030.

OGLE-2018-BLG-0677 presents a special case because the $3\,\sigma$
lower limit on its proper motion (derived from Figure~8 of
\citealt{ob180677}) is only $3.7\,\masyr$.  It is quite
plausible that the Einstein radius of this system is like the
others in Table~\ref{tab:6event}, i.e., $\theta_\e\sim 115\,\muas$, in
which case $\mu_\rel\sim 10\,\masyr$, implying that it would
be feasible to image this system in 2030, like the others.
This could be tried,
but a non-detection would not clearly establish that the host
was a BD.

Some progress is possible using present telescopes.  For example,
MOA-2011-BLG-262 is already separated by $\Delta\theta\sim 130\,\mas$,
and so it should be possible to probe companions to a contrast ratio
of $\sim 5$ magnitudes on Keck.  However, a non-detection would
yield only an upper limit on the host mass that would be well within
the stellar range. Note that \citet{mb11262} have presented a first
epoch for comparison.

If the lens is detected in these observations, then its mass can
be reliably inferred from the $K$-band flux, together with the improved
$\mu_\rel$ determination (and so improved $\theta_\e=\mu_\rel t_\e$
determination) from the lens-source separation measurement.
Hence, the planet mass can also be determined.  Non-detection of the
lens would imply that the host is a BD, and would also give an
upper limit on the mass of the planet, i.e., $m_p<q M_{\rm H-burn}$.
The relative fraction of BD and stellar hosts would constrain the
``mass function'', i.e., the mass function of low-mass stars and BDs that host
planets.  This could then be compared to the mass function of
(apparently) isolated stars and BDs, which can also be obtained
from microlensing.

Assuming that future AO observations confirm that the hosts of the
planets in Table~\ref{tab:6event} lie close to the star-BD
boundary, such objects host a wide variety of planets.  Adopting
$M_{\rm host}=M_{\rm H-burn}=0.075\,M_\odot$ for illustration, the six planets would
have (in order of mass) $m_p =[(2,12,51,135,300)M_\oplus,(7.8,8.8)M_J]$.
This distribution already hints at two populations of ``planetary''
companions of very low mass objects, genuine planets $m_p\la M_j$
formed by core-accretion and much more massive objects $m_p\gg M_J$,
drawn from the tail of BDs that are formed by gaseous collapse
in a manner similar to stars.

\subsection{{Patterns of Short-timescale Planetary Events}
\label{sec:pattern}}

There are several features of this sample that are important for
understanding the detectability of these systems.

One key feature is that in only one of these seven events
did followup observations play a role.  Indeed, in this case
(MOA-2011-BLG-262), follow-up observations (including auto-followup by MOA)
were essential in the interpretation of the anomaly.  The
remaining six cases were survey-only detections, and the KMTNet survey
(which began in 2015) was crucial in all six.

Another feature of Table~\ref{tab:6event} is that the planetary
signals for all seven events are generated by resonant or ``near-resonant''
caustics.  Six come from resonant caustics, i.e.,
the six-sided caustics formed by the ``merger'' of central and planetary
caustics that occurs as $s\rightarrow 1$.  One comes from a
``near-resonant'' caustic structure, which was defined by
\citet{ob190960} as topologically disjoint caustic structures
that have ridges (or valleys) of excess magnification of at least
10\% that connect the central and planetary caustics.  However, this
is not surprising.
\citet{ob190960} showed that the great majority of microlensing
planets are found in events from these two caustic topologies in roughly
the proportion 3:2.  Hence, from binomial statistics, the probability
that one or fewer from a sample of seven would be near-resonant is $p=16\%$.
Nevertheless, this feature is important in that it means
that these systems are detected from relatively short-lived anomalies
near the peak of relatively high-magnification events.

Finally, all seven events have faint source-star magnitudes, $I_s>19.2$.
This is mainly explained by the fact that faint sources are much
more common than bright ones.  However, it does emphasize that in typical
real cases, the source has only marginally brightened two days before
peak, and has hardly brightened one day before peak.  That is, for
a typical $t_\e\sim 4\,$day event on an $I_s=19.5$ source, the
``difference star'' on subtracted images is just $I_{\rm diff}=19.5$
two days before peak and $I_{\rm diff}=18.3$ one day before peak.
The OGLE EWS system \citep{ews1,ews2} rarely alerts on single-night
excursions at this level,
and the MOA system \citep{bond01} never does.
Moreover, OGLE alerts are usually issued about
10 hours after the end of the night.  This explains why there were
no such detections based on follow-up observations of OGLE alerts.

By contrast, MOA attempts to issue alerts shortly after a fast-rising
event is detected.  The MOA threshold of detection is much brighter
than OGLE, but for fast-rising events, this is more than compensated
by this quick response.  In the case of MOA-2011-BLG-262, MOA issued
its alert about 6 hr after the first observation of the night, and
just 50 min after three observations confirmed a rapid rise.  This
enabled the first followup observations less than 30 minutes later,
allowing full coverage of the anomaly.  Without this alert, there
would have been only one or two data points over the anomaly.
Nevertheless, this is truly a unique example from 14 years of the
MOA-II experiment.  MOA did not issue alerts for any of the other
events in Table~\ref{tab:6event}, except for KMT-2019-BLG-0371,
which it alerted at about the mid-point of the anomaly\footnote{Using online
MOA and OGLE data that covered only the caustic entrance of KMT-2019-BLG-0371,
Valerio Bozza issued an anomaly alert for this
event at UT 08:30 on 19 April, which gave a basically correct estimate
of the event parameters.  However, the anomaly had just ended at the time
this alert was issued.}.

MOA did issue an alert for
MOA-2015-BLG-337 on HJD$^\prime = 7214.02$, which would have been
plenty of time to initiate intensive observations from Chile
at HJD$^\prime \sim 7214.7$, which could have distinguished the
two models with $q$ differing by a factor $\sim 20$.  See
Figure~1 from \citet{mb15337}.  However, the main team that could
have carried out such observations, $\mu$FUN, had discontinued
intensive followup observations at this time in order to
focus on {\it Spitzer} microlensing candidates \citep{yee15}.  There were survey
observations from KMTC in Chile, but these commissioning-year data
were of insufficient quality.  We note that MOA-bin-29 \citep{moabin29}
was not discovered in real time, so there was no possibility of followup
observations during the 2006 season, and hence there were substantial
gaps in the light-curve coverage.  Moreover, it is not completely clear
that the ``Wide-1''/``Wide 3'' degeneracy (with different $q$ by a factor 2.7)
could have been resolved by additional coverage.

In brief, all six of the survey-only short-$t_\e$ planets in
Table~\ref{tab:6event} occurred after the start of KMTNet observations
in 2015, and KMTNet data were essential to all six.  During the nearly
two decades of microlensing planet detections, there has been only
one short-$t_\e$ planet detected by means of survey-plus-followup observations.
The above discussion shows that these patterns are reasonably well understood.

Thus, if the currently very small sample of these important systems
is to be increased, the most likely path is to improve the harvest
from KMT survey.

\subsection{{Path to Additional Short-$t_\e$ Planetary Events}
\label{sec:path}}

There are two obvious paths to finding more planetary anomalies
in archival short-$t_\e$ KMT events.  First, as noted by \citet{ob191053},
the online data reductions were substantially improved starting in 2018.
Simply applying the same algorithms to 2016 and 2017 data would make
it much easier to spot anomalies by eye, or to find them by the
automated technique described by \citet{ob191053}.  We note that of
the six survey-only detections in Table~\ref{tab:6event}, three were
from prior to 2018.  Of these three, one was not discovered by KMT
(OGLE-2015-BLG-1771), one was part of the special 2016 search and so
was reduced using the new algorithm (KMT-2016-BLG-2605), and one was
a massive planet with a huge, easily discernible anomaly (KMT-2016-BLG-1820).
Hence, updating the 2016-2017 reductions, which is currently underway,
may well increase the detectability of moderate mass-ratio planets
for these seasons\footnote{It is not clear that it will be possible
to improve the pipeline light curves for 2015, due to the lower
quality of this commissioning-year data.  There are currently no plans
to do so.}.

A second path would require a small alteration of the program outlined
by \citet{ob190960} to make TLC reductions for all ``high-magnification''
events, defined as perhaps $A_\max > 20$ or $A_\max > 10$.  Subtle anomalies,
like the one seen in OGLE-2018-BLG-0677 \citep{ob180677}, will only
appear convincing (or may only be noticed) in high-quality TLC reductions.
Subtle anomalies may reflect very low-mass planets (as in that case),
or somewhat higher mass planets in events for which the source passes
farther from the caustics.  Excluding the two high-$q$ events
(KMT-2019-BLG-0371 and KMT-2016-BLG-1820), whose pronounced anomalies
are easily recognizable without TLC reductions, the remaining survey-only
events have peak magnifications (as judged by $A_\max = 1/u_0$) of
$A_\max = (9,10,20,29)$.  And machine PSPL fits could easily underestimate
the peak magnification, depending on how these fits were affected by
the anomaly.  Therefore, the $A_\max$ criterion for TLC reductions
could be loosened for short-$t_\e$ events.


The problems posed by anomaly-dominated events like KMT-2019-BLG-2605
are more challenging.  While this event constitutes only 14\% of the
current sample and may therefore appear relatively inconsequential,
it arrived in the sample by a quite accidental route.  Hence, it
could be under-represented.  It would be impractical to repeat
the EventFinder searches of archival KMT data, but going forward,
the human reviews of the machine-selected EventFinder and
AlertFinder \citep{alertfinder} candidates could be more aggressive
for short events.  In particular, when there are magnified data from
only one night for each observatory and the event is anomalous,
the machine alignment of the data can be radically incorrect, and
one or more data sets can even be eliminated from the fit.  Recognition
of these issues could enable more potentially-anomalous, short events
to be conditionally selected at this stage.

For the same reason, it is possible that anomalous EventFinder events that
have been selected are being overlooked in manual reviews of the
KMTNet webpage.  That is, the poor machine alignment of the different
data sets can make the event look like ``not microlensing'', leading to
it not being selected for further analysis.  Simple recognition of
this possibility, based on the experience of KMT-2016-BLG-2605,
may lead to a revised preliminary assessment of such events.

Here, it should be pointed out that archival events are, in some
sense, more productive than prospective ones, because they
will become eligible for AO imaging sooner.

\acknowledgments
%
%
%
This research has made use of the KMTNet system operated by the Korea
Astronomy and Space Science Institute (KASI) and the data were obtained at
three host sites of CTIO in Chile, SAAO in South Africa, and SSO in
Australia.
Work by C.H. was supported by the grants of National Research Foundation
of Korea (2020R1A4A2002885 and 2019R1A2C2085965).
%
%
%
%

\appendix
\section{Assessment of Source Color}
\label{sec:append}

\begin{figure}
\plotone{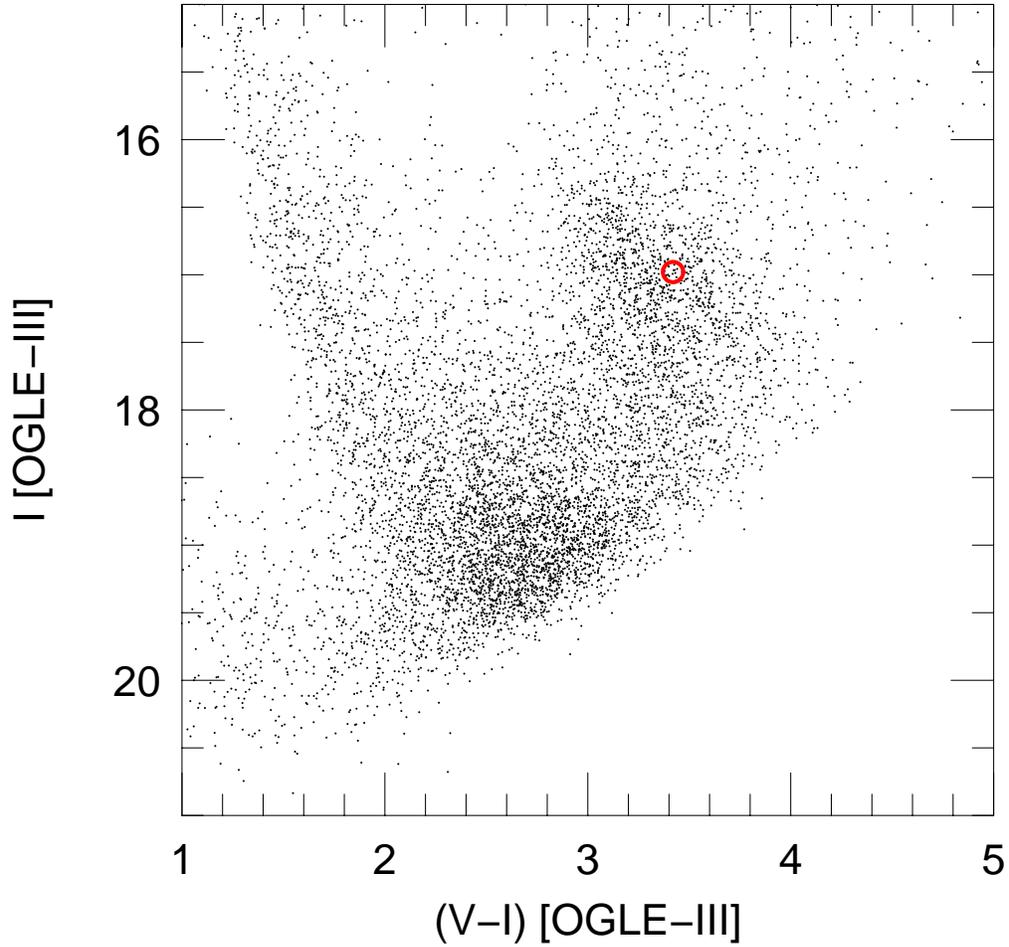}
\caption{OGLE-III \citep{oiiicat} CMD for stars within $180^{\prime\prime}$ of
KMT-2016-BLG-2605.  The clump is clearly visible, but is extended from
upper left to lower right, indicating strong differential reddening.  The
red circle is the clump center as determined from the $60^{\prime\prime}$ CMD in
Figure~\ref{fig:ocmd60}, which is clearly not aligned with the center of the
clump ``feature'' in this diagram.
}
\label{fig:ocmd180}
\end{figure}

\begin{figure}
\plotone{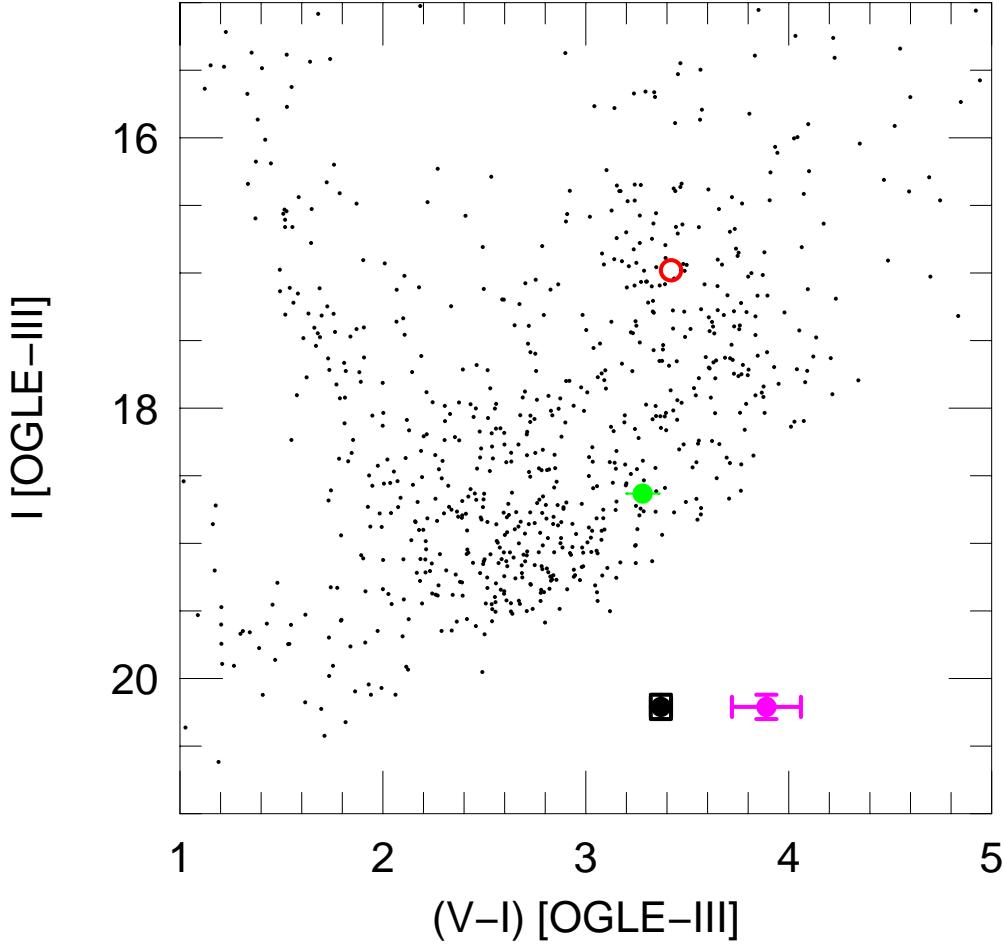}
\caption{OGLE-III CMD for stars within $60^{\prime\prime}$ of
KMT-2016-BLG-2605.  The clump is less clearly visible than in
Figure~\ref{fig:ocmd180} but with the aid of that figure it can be identified.
The red circle is the clump centroid.  The magenta circle represents the
CMD position of the source as determined from the light curve alone.
The black circle is the adopted CMD source position after incorporating
information from the {\it HST} CMD shown in Figure~\ref{fig:hstcmd}.
The implications of these two different source positions and how they
can eventually be distinguished is discussed in the Appendix.
The green circle represents the baseline-object position $[(V-I),I]_{\rm base}$,
where $I_{\rm base} = 18.63$ comes directly from the OGLE-III catalog
\citep{oiiicat} and
$(V-I)_{\rm base} = 3.28$ is derived by combining $I_{\rm base}$ with
$K_{\rm base} = 14.89\pm 0.07$ from the VVV catalog \citep{vvvcat}, and then
transforming from $(I-K)$ to $(V-I)$ using matched stars between OGLE-III and
VVV.  The $K$ magnitude of the baseline object can be helpful in future AO
imaging.  See the Appendix.
}
\label{fig:ocmd60}
\end{figure}

Figure~\ref{fig:ocmd180} shows OGLE-III \citep{oiiicat} stars
within a $180^{\prime\prime}$ circle, centered on the lensing event.
The clump is easily visible, but it is extended upper-left to
lower-right, which is a standard signature of differential
reddening.  Hence, we should be cautious about identifying
the centroid of the clump feature in this diagram with the
center of the clump at the position of the event.
Figure~\ref{fig:ocmd60} shows OGLE-III stars in a $60^{\prime\prime}$
circle centered on the event.  The clump is less visible,
but guided by Figure~\ref{fig:ocmd180}, it can be recognized,
and its center is marked by a red circle,
$[(V-I),I]_{\rm cl}= (3.42,16.98)\pm (0.02,0.04)$.  This same position
is marked by a circle in Figure~\ref{fig:ocmd180}, which demonstrates
that the centroid of the clump feature has indeed shifted fainter and
redder from the first to the second figure.

The source magnitude is well measured from the microlensing fit
in the KMTS pyDIA system, $I_{s,{\rm kmts}} = 20.06 \pm 0.04$. By comparing
field-star photometry from OGLE-III with that of the KMTS pyDIA reductions, we
find $I_{\rm kmts} - I_{\rm ogle-iii} =-0.15\pm 0.01$, implying
$I_{s,{\rm ogle-iii}} = 20.21 \pm 0.05$.
Hence, the offset in brightness of the source relative to the clump is
\begin{equation}
\Delta I = I_s - I_{\rm cl} = 3.23 \pm 0.07 .
\label{eqn:offset}
\end{equation}

However, the color offset $\Delta(V-I)$ is substantially more difficult to
determine.  There is only one substantially magnified $V$-band data point.
This would make it difficult to measure the source color under any
circumstances because there would be no internal check on the measurement.
In addition, as we report below, the image quality of the one magnified
$V$-band point exhibits some problems.  We therefore begin by asking
what can be deduced about the source color without a measurement
from the light curve.

\begin{figure}
\plotone{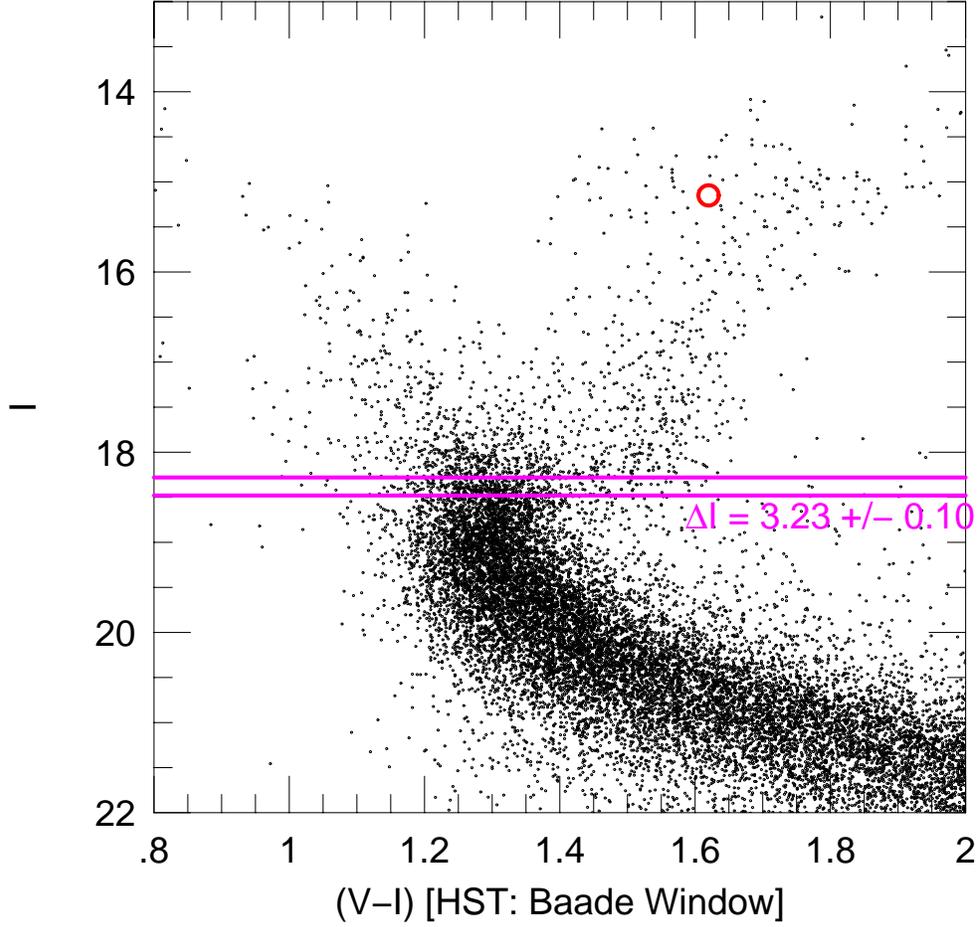}
\caption{{\it HST} CMD from the observations of the Baade Window by
\citep{holtzman98}.  The clump centroid, marked by a red circle,
is at $[(V-I),I]_{\rm cl} = (1.62,15.15)$ \citep{mb07192}.  The magenta
lines, lying $\Delta I=3.23\pm 0.10$ below the clump represent the
$1.5\,\sigma$ range for the source brightness relative to the clump
The source color, based on a single KMTS $V$-band measurement is
$\Delta(V-I) = 0.37\pm 0.17$ redward of the clump.  So, on this
diagram it would be at $(V-I)_{HST,\rm BW} = 1.99\pm 0.17$, i.e.,
at the extreme right of the magenta band.  However, it is not
shown to avoid clutter.  See the Appendix for why this
is most likely due to a large statistical fluctuation.  The
adopted color offset is $\Delta(V-I)_s= -0.06\pm 0.05$, consistent with
the subgiants at $(V-I)_{HST,\rm BW} = 1.56\pm 0.05$ in this diagram.
See black circle in Figure~\ref{fig:ocmd60}
}
\label{fig:hstcmd}
\end{figure}

Figure~\ref{fig:hstcmd} shows the {\it Hubble Space Telescope (HST)}
CMD for a Baade Window field constructed by \citet{holtzman98}.
The red circle shows the clump centroid $[(V-I),I]_{\rm BW} = (1.62,15.15)$
as determined by \citet{mb07192}.  The two magenta lines are displaced
$\pm 0.1$ (i.e., $1.5\,\sigma$) from the best estimate of the offset
(Equation~(\ref{eqn:offset})) for KMT-2016-BLG-2605, $\Delta I=3.23$.
Based on the stars between these two lines, we can make three
different characterizations of the stars at this offset:
\begin{equation}
-0.47 < \Delta(V-I) < -0.01;
\quad
 \langle\Delta(V-I)\rangle = -0.31\pm 0.09;
\quad
 \Delta(V-I)_{\rm median} = -0.32_{-0.07}^{+0.08} .
\label{eqn:hstdvmi}
\end{equation}
The first is the full ``reasonably populated''  region of the strip.
The second is the mean and standard deviation of this populated
region.  The third is the median and (16,84)th percentiles of the
full distribution.  If there were absolutely no other information
about the source color, one would take either the mean or median
estimator, which in the present case are almost identical.

We will next consider the color measurement based on the light curve,
i.e., on the single magnified $V$-band measurement.
However, before proceeding,
we should ``predict'' the KMTS $V$-band flux
measurement at HJD$^\prime = 7565.4446$ based on the contemporaneous
$I$-band flux measurement ($F_I = 35688 \pm 325$), and the range
of ``reasonably populated'' $\Delta(V-I)$ given by
Equation~(\ref{eqn:hstdvmi}).  To do so, we take note of the
offset (measured from field stars)
$(V-I)_{\rm kmts} - (V-I)_{\rm ogle-iii} = 0.27\pm 0.02$, the KMTS instrumental
photometric zero points $(V_{\rm zero}=28.65$ and $I_{\rm zero} =28.00)$,
and the OGLE-III clump centroid $(V-I)_{\rm cl}=3.42\pm 0.02$.
That is,
\begin{equation}
F_{V,\rm predicted} = 0.0608 F_I \times 10^{-0.4\Delta(V-I)} \rightarrow
2170\times 10^{-0.4\Delta(V-I)},
\label{eqn:vpredict}
\end{equation}
and hence, for the
full ``reasonable range'' of $\Delta(V-I)$, we predict,
$2200 < F_{V,\rm predicted} < 3360$,
which should be compared to the observed $V$-band difference flux
returned by the photometry program, $F_V= 1545\pm 238$.  That is,
the observed flux lies $2.75\, \sigma$ below the ``reasonable range''.
This could mean that the source is a very rare, exceptionally red
star, that the error bar has been substantially underestimated,
or that the measured value is the outcome of a rare statistical
fluctuation.

We find no evidence that the photometry program has generally
underestimated the error bars on the $V$-band light-curve measurements.
In particular,
we look at the distribution of $\sigma_i/F_i$ of the 87
measurements apart from the well-magnified one and the one
on the previous night at modest magnification (for which the predicted
difference flux is $<1\,\sigma$).  For these 87,
the expected difference flux is zero
to high precision.  We find that this distribution is consistent with a Gaussian
of zero mean and unit variance.

We examine the original and subtracted images for the magnified point
and compare these to several images for unmagnified points.  In the original
images, the source generally appears isolated, and there are only
very faint stars within a few arcseconds.  With the exception of
the magnified point, the subtracted images generally appear ``blank'' at,
and for several arcseconds around, the source.  Hence, there is no
obvious cause for difficulty in performing the photometry, in
line with the fact (just reported) that the normalized error distribution
is a unit Gaussian.

The magnified image is taken seven days after passage of the full moon through
the bulge, so that the background is about 2.8 times the dark-time
level.  As a result of this higher background, the subtracted
image appears substantially more mottled than for dark-time images.
Nevertheless, the background level (453 ADU per pixel) is by
no means high.  Similarly, the seeing has a FWHM$_{\rm see}\sim 2.75^{\prime\prime}$,
which is higher than the median ($2.39^{\prime\prime}$), but hardly
unusual (66th percentile).  And also similarly, the transparency is
about 88\% relative to typical good nights, which is hardly out of
the normal range.

Finally, we consider the general possibility that the program has
underestimated the error bar for some ``unknown reason''.  The program
makes its estimate by varying the fit parameters and finding the
change of $\chi^2$ that results.  This should be robust, but for
any relatively complex program, one can imagine that it confronts
some unexpected condition and makes a catastrophic error.  As
a sanity check, we make a naive estimate of the error as being
proportional to [FWHM$_{\rm see}\times$(FWHM$_{\rm back}$/transparency)$^{1/2}$],
where FWHM$_{\rm back}$ is the full width at half maximum of the difference-flux
pixel-count distribution of the subtracted image.  For images that are well
below sky, this scaling should be close to accurate.   We normalize this
estimator to an image with low background (164),
good seeing ($1.51^{\prime\prime}$) and 100\% relative transparency
and find only a 19\% difference in predicted versus reported error bars.
This is an order of magnitude below what would be required to explain
the apparent discrepancy (and also goes in the wrong direction).

In brief, the source location is isolated,
the program overall evaluates the $V$-band errors correctly,
the seeing and background of the magnified image
are slightly worse than average but by no means unusual, and
a simple sanity check confirms the program's evaluation of the
error bar.

If the $2.75\,\sigma$ discrepancy between prior expectations and the observed
data point are to be explained within the context of Gaussian
statistics, then $p_{\rm gauss}=0.0031$.  Therefore, before accepting this
explanation, we should consider various others that are of such
low probability that they would normally be dismissed without
detailed investigation.

First, the source may actually be drawn from the extremely red
population that is the reflected in the {\it HST} CMD.  Of the
414 stars shown between the magenta lines, two are within the
$1\,\sigma$ range of the magnified point,
$\Delta(V-I) = +0.37\pm $0.17, and one other
is redward of this range.  This fraction, $ 3/414 = 0.007$, is greater
than $p_{\rm gauss}$, and so this possibility
should be considered.  However, from the morphology
of Figure~\ref{fig:hstcmd}, these very red stars appear to be part of
the disk red-dwarf population that lives ``above'' the bulge
main sequence in this diagram.  As such, the red stars within
the magenta bands lie about 3 mag in front of the bulge in distance
modulus, i.e., at about $D_S = 2\,\kpc$.  In addition to being extremely
rare (as just noted), the optical depth to microlensing of such nearby
disk sources is two orders of magnitude lower than for bulge sources.
Thus, we regard this potential explanation as highly improbable.

\citet{bensby17} provide some corroboration of this assessment.
They obtained 91 high-resolution spectra of highly-magnified
``dwarf and subgiant'' sources.  These were almost all selected solely
on source-brightness relative to the clump
(i.e., not giants) and observability (magnified enough
to obtain a good spectrum), which in practice essentially produced
an unbiased sample of turnoff stars and subgiants.  None of these 91
had spectroscopic temperatures cooler than the clump ($\sim 4750\,$K).
See upper panel of their Figure~7.  While 0/91 does not place
restrictions at the level of $p_{\rm gauss}$, it does demonstrate
that such extremely red microlensed sources lying $3\,$mag below
the clump are very rare.

Another possibility is that the microlensing model is incorrect,
so that the source is actually brighter (relative to the clump)
than the magenta band.  For example, the source is 0.8 mag brighter
for Local 3 than for Local 1.  However, the {\it HST} CMD is even
less populated 0.8 mag above the red end of the magenta band than
in the band itself.  One might posit that there is another solution
with an even brighter source that we failed to discover.  However,
the source cannot be much brighter\footnote{It could be slightly
brighter because the source might be projected against a ``hole''
in the mottled background due to unresolved field stars \citep{mb03037}.
However, this effect is far too small to be relevant here.}
than the baseline object,
which is only $-2.5\log(1 + f_B/f_S)= -1.5\,$mag brighter than the
magenta bar.  This is still far below the region of the CMD
that is populated by upper giant-branch stars.

Yet another possibility is that the source is actually a giant in the
far side of the disk.  There would be extremely few such stars
in the {\it HST} CMD because it lies in the Baade Window at $b\sim -4$,
so that the line of sight intersects the bulge about
$z_{\rm bulge}\sim -550\,$pc from
the Galactic plane.  Far-disk sources are more plausible for
KMT-2016-BLG-2605 for which $b=-1.6$, so that $z_{\rm bulge}\sim -210\,$pc.
For example, at $D_S=12\,\kpc$, the line of sight passes
$z_{\rm 12\,\kpc}\sim -320\,$pc from the plane, where potential source
stars remain plentiful.  Nevertheless, in order to access the red
upper-giant-branch stars, the source would have to have a distance
modulus at least 2.5  larger than the bulge, i.e.,
$D_S>25\,\kpc$ or $\sim 2 R_0$ from the Galactic center on the far
side of the Galaxy, with $z_{\rm 25\,\kpc}\sim -700\,$pc from the plane.
This is a very thinly populated region of the Galaxy.  While we do
not exclude this possibility, and we report its implications further
below, we consider it less likely than a statistical error in
the $V$-band measurement.

We conclude that the most plausible resolution is that the
source color is toward the red end of the ``reasonable range''
from Equation~(\ref{eqn:hstdvmi}) and that the very red
light-curve measurement is the result of a relatively large
statistical fluctuation.  We therefore adopt
\begin{equation}
\Delta(V-I)_s = -0.06\pm 0.05
\Rightarrow (V-I)_{s,0} = 1.00\pm 0.05
\label{eqn:vmi-adopt}
\end{equation}
However, we also consider the possibility that the light-curve measurement is
actually correct (due, e.g., to a very distant far-disk source),
i.e., $(V-I)_0 = 1.43\pm 0.17$, and thus we trace the consequences
of this possibility.

\subsection{Effects of Alternate Color Estimate}
\label{sec:altcol}

We have adopted a source color $(V-I)_s=3.37\pm 0.05$
(equivalently, $(V-I)_{s,0}=1.00\pm 0.05$) by combining prior information
from the \Citet{holtzman98} CMD with the KMTS color measurement.
Here, we consider the consequences if the source color is actually
given by the KMTS measurement, i.e., $(V-I)_s=3.89\pm 0.17$
(equivalently, $(V-I)_{s,0}=1.43\pm 0.17$).

The first point is that the true source color can eventually be determined
by high-resolution imaging, and indeed this may already be possible
with 10m-telescope class AO imaging.  Using \citet{bb88} to convert from
$(V-I)$ to $(I-K)$, and adopting $E(I-K) = 2.35$ from
Section~\ref{sec:identify},
we find $K_s =16.48\pm 0.08 $ or $K_s = 15.96_{-0.19}^{+0.35}$,
for the two scenarios.  These values
can be compared to the $K$-band magnitude of the baseline object from
the VVV survey \citep{vvvcat} of $K_{\rm base} = 14.89\pm 0.07$.  That is,
roughly 23\% or 37\% of the baseline-object $K$-band light comes from the
source.

There are logically only four possibilities for the remainder of the
$K$-band light: the lens, a companion to the lens, a companion to the
source, or an ambient star (or some combination).  It is very unlikely that
an ambient star would lie within the $\sim 55\,\mas$ point-spread function (PSF)
of a 10m telescope.  If the blended light were due to the lens or a companion
to the lens, then by 2021, it would have already separated from the
source by $\Delta\theta = \mu_\rel\Delta t = 61\pm 5\,\mas$
(or $79\pm 7\,\mas$).  In either case, the source and lens could
be separately resolved.  See Figure~1 of \citet{ob05071c} for a
separate resolution of a source and lens with flux ratio 3.15
at $\Delta\theta = 55\,\mas$, and Figure~1 of \citet{ob120950b} for
an unambiguous distinction between a source and lens with flux ratio 1.46
at $\Delta\theta = 34\,\mas$, both based on $K$-band observations with the
Keck telescope.  Thus, unless the blended light is due to a companion
to the source (which would then be a lower-giant-branch star, which
is a priori unlikely due to its short lifetime),
the source color could almost certainly be determined by observations in 2021.

Such immediate observations might also resolve the lens, and, even if not,
would give a definite prediction as to when the lens could be resolved.
For example, suppose that these observations found that $K_s=15.96$
(with small error).
One could then conclude that $\mu_\rel=15.8\pm 0.8\,\masyr$ so that
the annulus of possible lens positions (at 1.5 FWHM)
could be predicted with precision.  Hence, one could already detect the
lens, or place strong constraints on its brightness.  If the source
proved to be substantially fainter in $K$, this would imply a smaller
$\theta_\e$ and hence a smaller $\mu_\rel$.  However, it would still
be possible to use this information to predict when the lens would be
observable.  As discussed in Section~\ref{sec:ensemble} non-detection
of the lens in relatively shallow imaging would indicate the need
for deeper imaging, either on 10m or future 30m class telescopes.



%

\begin{deluxetable}{lccccc}
\tablecolumns{6} \tablewidth{0pc} \rotate \tablecaption{\textsc{Best
Solutions without Flux Constraint}} \tablehead{\colhead{Parameters}
& \colhead{Local 1} & \colhead{Local 1a} & \colhead{Local 1b} &
\colhead{Local 2} & \colhead{Local 3}} \startdata
  $\chi^2/\rm{dof}$             &3596.308/3598        &3603.450/3598        &3605.222/3598        &3606.825/3598        &3606.218/3598       \\
  $t_0-2457560$                 &5.451 $\pm$ 0.021    &5.457 $\pm$ 0.010    &5.473 $\pm$ 0.010    &5.643 $\pm$ 0.023    &5.473 $\pm$ 0.013 \\
  $u_0$                         &0.049 $\pm$ 0.004    &0.046 $\pm$ 0.004    &0.047 $\pm$ 0.004    &0.097 $\pm$ 0.006    &0.085 $\pm$ 0.007    \\
  $t_{\rm E}$ $(\rm{days})$     &3.370 $\pm$ 0.139    &3.291 $\pm$ 0.123    &3.023 $\pm$ 0.146    &2.237 $\pm$ 0.057    &2.319 $\pm$ 0.104   \\
  $s$                           &0.939 $\pm$ 0.011    &0.924 $\pm$ 0.006    &0.914 $\pm$ 0.004    &1.827 $\pm$ 0.064    &0.797 $\pm$ 0.014     \\
  $q$                           &0.012 $\pm$ 0.002    &0.013 $\pm$ 0.002    &0.007 $\pm$ 0.001    &0.242 $\pm$ 0.074    &0.019 $\pm$ 0.003     \\
  $\alpha$ $(\rm{rad})$         &0.101 $\pm$ 0.019    &0.180 $\pm$ 0.018    &0.057 $\pm$ 0.031    &2.580 $\pm$ 0.040    &-0.042 $\pm$ 0.065     \\
  $\rho$ $(10^{-2})$            &1.203 $\pm$ 0.125    &1.068 $\pm$ 0.079    &1.145 $\pm$ 0.156    &3.345 $\pm$ 0.262    &1.651 $\pm$ 0.269     \\
  $f_S$ [KMTC]                  &0.151 $\pm$ 0.016    &0.163 $\pm$ 0.011    &0.206 $\pm$ 0.024    &0.320 $\pm$ 0.023    &0.336 $\pm$ 0.032    \\
  $f_B$ [KMTC]                  &0.388 $\pm$ 0.016    &0.376 $\pm$ 0.011    &0.332 $\pm$ 0.024    &0.220 $\pm$ 0.023    &0.203 $\pm$ 0.032     \\
  $f_S$ [KMTS]                  &0.151 $\pm$ 0.010    &0.141 $\pm$ 0.009    &0.170 $\pm$ 0.014    &0.389 $\pm$ 0.023    &0.314 $\pm$ 0.026     \\
  $f_B$ [KMTS]                  &0.436 $\pm$ 0.010    &0.446 $\pm$ 0.009    &0.417 $\pm$ 0.014    &0.198 $\pm$ 0.023    &0.273 $\pm$ 0.026     \\
  $f_S$ [KMTA]                  &0.096 $\pm$ 0.006    &0.090 $\pm$ 0.005    &0.103 $\pm$ 0.009    &0.234 $\pm$ 0.013    &0.200 $\pm$ 0.016     \\
  $f_B$ [KMTA]                  &0.270 $\pm$ 0.006    &0.276 $\pm$ 0.005    &0.263 $\pm$ 0.009    &0.133 $\pm$ 0.013    &0.167 $\pm$ 0.016     \\
  $I_{S,{\rm kmts}}-I_{S,{\rm kmtc}}$ &-0.002 $\pm$ 0.066    &0.152 $\pm$ 0.062    &0.208 $\pm$ 0.048    &-0.213 $\pm$ 0.049   &0.072 $\pm$ 0.045   \\
  $t_*$ $(\rm{days})$           &0.041 $\pm$ 0.004    &0.035 $\pm$ 0.002    &0.035 $\pm$ 0.003    &0.075 $\pm$ 0.005    &0.038 $\pm$ 0.005     \\
\enddata
\label{tab:local_1-3}
\end{deluxetable}

\begin{deluxetable}{lccc}
\tablecolumns{4} \tablewidth{0pc} \tablecaption{\textsc{Best
Solutions with Flux Constraint}} \tablehead{\colhead{Parameters} &
\colhead{Local 1} & \colhead{Local 2} & \colhead{Local 3}}
\startdata
  $\chi^2/\rm{dof}$             &3597.963/3599        &3622.808/3599        &3613.757/3599       \\
  $t_0-2457560$                 &5.451 $\pm$ 0.005    &5.553 $\pm$ 0.039    &5.500 $\pm$ 0.008 \\
  $u_0$                         &0.049 $\pm$ 0.004    &0.091 $\pm$ 0.012    &0.081 $\pm$ 0.007    \\
  $t_{\rm E}$ $(\rm{days})$     &3.405 $\pm$ 0.128    &2.207 $\pm$ 0.071    &2.402 $\pm$ 0.100   \\
  $s$                           &0.940 $\pm$ 0.005    &1.756 $\pm$ 0.093    &0.787 $\pm$ 0.008     \\
  $q$                           &0.012 $\pm$ 0.001    &0.175 $\pm$ 0.092    &0.023 $\pm$ 0.002     \\
  $\alpha$ $(\rm{rad})$         &0.104 $\pm$ 0.010    &2.483 $\pm$ 0.097    &0.053 $\pm$ 0.043     \\
  $\rho$ $(10^{-2})$            &1.192 $\pm$ 0.083    &3.683 $\pm$ 0.283    &1.298 $\pm$ 0.203     \\
  $f_S$ [KMTC]                  &0.145 $\pm$ 0.009    &0.354 $\pm$ 0.031    &0.291 $\pm$ 0.025    \\
  $f_B$ [KMTC]                  &0.393 $\pm$ 0.009    &0.186 $\pm$ 0.031    &0.248 $\pm$ 0.025     \\
  $f_S$ [KMTS]                  &0.150 $\pm$ 0.009    &0.364 $\pm$ 0.032    &0.300 $\pm$ 0.025     \\
  $f_B$ [KMTS]                  &0.437 $\pm$ 0.009    &0.223 $\pm$ 0.032    &0.287 $\pm$ 0.025     \\
  $f_S$ [KMTA]                  &0.095 $\pm$ 0.006    &0.223 $\pm$ 0.018    &0.190 $\pm$ 0.016     \\
  $f_B$ [KMTA]                  &0.271 $\pm$ 0.006    &0.144 $\pm$ 0.018    &0.176 $\pm$ 0.016     \\
  $I_{S,{\rm kmts}}-I_{S,{\rm kmtc}}$       &-0.033 $\pm$ 0.005   &-0.032 $\pm$ 0.005   &-0.033 $\pm$ 0.005   \\
  $t_*$ $(\rm{days})$           &0.041 $\pm$ 0.002    &0.081 $\pm$ 0.005    &0.031 $\pm$ 0.004     \\
\enddata
\label{tab:const_1-3}
\end{deluxetable}

 \begin{deluxetable}{lrrrrrrrl}
 \tablecolumns{10} \tablewidth{0pc}
 \tablecaption{\textsc{7 Planetary Events with $t_\e<7\,$days}}
 \tablehead{\colhead{Event} & 
\colhead{$t_\e$} &
\colhead{$q$} &
\colhead{$\ln s$} &
\colhead{$\theta_*$} &
\colhead{$\theta_\e$} &
\colhead{$\mu_\rel$} &
\colhead{$I_s$} & 
\colhead{Caustic Type} }
 \startdata
KMT-2016-BLG-2605 &3.41&0.0120&$-0.06$   &1.38 & 116 & 12.3 & 20.21 & Resonant\\
MOA-2011-BLG-262  &3.87&0.00047&$\pm 0.05$&0.78& 136 & 12.9 & 19.34 & Resonant\\
OGLE-2015-BLG-1771&4.28&0.00538&0, +0.18  &0.49& 111 & 9.5  & 21.77 & Resonant\\
KMT-2018-BLG-0748 &4.38&0.00203&$-0.06$   &1.21& 111 & 9.2  & 19.21 & Resonant\\
KMT-2016-BLG-1820 &4.81&0.11300&$+0.15$   &0.81& 123 & 9.3  & 19.38 & Resonant\\
OGLE-2018-BLG-0677&4.94&0.00008&$-0.09$,$-0.02$&0.79&$>49$ &$>3.6$&19.32& Near-Resonant\\
KMT-2019-BLG-0371 &6.53&0.08,0.12&$-0.19$,$+0.45$&0.92& 135 & 7.6  & 19.76 & Resonant\\
 \enddata
 \tablecomments{$t_\e$ is in days, $\theta_*$ and $\theta_\e$ are in $\muas$,
and $\mu_\rel$ is in $\masyr$. }
 \label{tab:6event}
 \end{deluxetable}

\begin{deluxetable}{lc}
\tablecolumns{2} \tablewidth{0pc} \tablecaption{\textsc{Physical
parameters}} \tablehead{\colhead{Quantity} & \colhead{Local 1} }
\startdata
  $M_{\rm host}$ $[M_\odot]$       &$0.064_{-0.033}^{+0.099}$\\
  $M_{\rm planet}$ $[M_J]$        &$0.771_{-0.401}^{+1.183}$\\
  $a_{\bot}$ [au]                 &$0.681_{-0.110}^{+0.102}$\\
  ${\it D_L}$ [kpc]               &$6.421_{-0.991}^{+0.878}$\\
 \enddata
\label{tab:bayes}
\end{deluxetable}

\end{document}